\documentclass[aps,prb,twocolumn,showpacs,superscriptaddress]{revtex4}
\usepackage{graphicx}
\usepackage{dcolumn}

\begin{document}


\title{Magnetism in a TiO$_2$/LaAlO$_3$ heterostructure: an ab initio study about  the role of oxygen vacancies}



\author{Valeria Ferrari and Mariana Weissmann}
\affiliation{Departamento de F\'{\i}sica, Comisi\'on Nacional de 
Energ\'{\i}a At\'omica, Gral. Paz 1499, 1650 San Mart\'{\i}n, Buenos Aires, Argentina.}

\date{\today}

\begin{abstract}

In this work we study the electronic structure and magnetism of a TiO$_2$ film grown on another non-magnetic oxide such as a LaAlO$_3$ (001) substrate, concentrating on the role played by structural relaxation and oxygen vacancies. Using Density Functional Theory ab-initio methods, we study the free-standing anatase film as well as the interfaces with either the LaO or AlO$_2$ planes of LaAlO$_3$, focusing on the possibility of magnetic solutions.
Our results show that the interface LaO/TiO$_2$ is favored against the AlO$_2$/TiO$_2$ one if no oxygen vacancies are present in the interface whereas the contrary happens when there are oxygen vacancies. In both cases, the cohesive energy is of the same order of magnitude but only the AlO$_2$/TiO$_2$ interface presents an stable magnetic solution.

\end{abstract}

\maketitle

\section{Introduction}

The development of modern microelectronics devices requires the knowledge of 
the electronic structure near interfaces. In this respect, oxide materials 
combine many of the important properties of semiconductors and add novel 
phases such as metal-insulator transitions \cite{CenNMat08}, superconductivity 
\cite{SupcondSTOLAO}, quantum Hall effect \cite{QuantumHalloxides}, orbital reconstruction \cite{YBCO_LCMO}, colossal 
magnetoresistance \cite{BrinkmanNatMat07},etc. One interesting phenomena reported recently is 
the change in the properties of complex oxides when they are arranged in a 
heterostructure. For example, LaAlO$_3$ (LAO) and SrTiO$_3$ (STO) are both insulators and non-magnetic 
in their bulk form but one possible interface between them is conductive \cite{OhtomoNature04} and 
ferromagnetic \cite{BrinkmanNatMat07,TsymbalJAP08}. The origin  of the ferromagnetism and 
the conductivity is not known, it has been attributed to the polar 
discontinuity between the charged layer (LaO) in contact with the neutral layer (TiO$_2$), that would produce 
a transfer of electrons at the interface, and also 
to the presence of  oxygen vacancies created  during film growth \cite{NakagawaNatMat06,FreemanPRB06,BrinkmanReview08}.

The previous findings may be related to the recent intense search for ferromagnetism 
above room temperature in dilute magnetic semiconductors and insulators.  The fabrication of these materials 
offers exciting possibilities for spintronic devices and the use of wide 
gap oxides is appealing for magnetooptical devices.  In that direction, 
room-temperature ferromagnetism has been observed in both doped 
\cite{ChambersReview06,PanAPL06}  and  
undoped insulating oxide thin films such as TiO$_2$, 
ZnO and HfO$_2$ \cite{yoonJMMM07,yoonJMMM07_addendum,sudakarJMMM08} grown over other oxides or semiconductors. This 
magnetic order can be  weak and the determination of its intrinsic 
character may require complementary, independent and careful techniques 
to be probed \cite{KasparPRB08,ChambersReview06,ClaudiaAPL08}.
Many authors have attributed an important role to  oxygen vacancies    
\cite{hongPRB06} and other defects \cite{FMandDefectsTiO2,EsquinaziAPL08} to produce this
magnetic ordering, but the effect of the substrate or the interface has 
not been taken into account in the interpretations.

Considering the open questions in the aforementioned systems, in this work
 we  explore the possibility of observing magnetism in an 
heterostructure formed by a simple oxide such a TiO$_2$ anatase and a LAO 
substrate. Experiments show that large terraces are formed on the substrate
surface, and that the growth is epitaxial in the (001) direction \cite{Lotnyk}.
However, it is nor clear if the surface is mostly composed of LaO planes
or of AlO$_2$ planes \cite{Francis,Lanier}. XPS experiments show that there
are Ti$^{+3}$ and Ti$^{+2}$ ions in these films, which may be due to the presence
of oxygen vacancies \cite{sasahara05} or to the polar catastrophe. The magnitude of the magnetization
has been found  to be proportional the vacancy concentration in some experiments, for
example in ref. \cite{PanAPL06}
We therefore concentrate on the effect of oxygen vacancies in the electronic 
structure and on the interfacial structural relaxation, as it is 
expected that the electric fields due to the dipole layer at the interface 
may lead to important changes in the positions of the atoms with respect 
to the bulk materials, as it happens in other interfaces 
\cite{NakagawaNatMat06}.  For this purpose we perform ab-initio calculations for the different 
interfaces present in the TiO$_2$/LAO heterostructure using computational codes such as SIESTA \cite{SIESTAmethod} or Wien2k \cite{Wien2k} and also
different unit cells.

We first study separately the  component systems:  TiO$_2$ and  LAO in bulk 
and as free standing slabs, representing the deposited anatase film and the 
substrate, respectively. As the ab-initio codes allow only the study of three 
dimensional periodic systems, the films are represented by repeated slabs with 
enough free space between them. For the heterostructure, two different types 
of unit cell were used: trilayer slabs LAO/TiO$_2$/LAO and superlattices 
...TiO$_2$/LAO/TiO$_2$/LAO...

The paper is organized as follows: In Sec.  \ref{calculationaldetails} we very briefly outline our 
calculational approach. In Sec.\ref{bulkandsurface} we discuss the bulk and surface electronic
properties of the component materials. In Sec. \ref{superlattices} we present the results
for the trilayers and superlattices and  in Sec. \ref{conclusions} our conclusions.

\section{Calculation details }
\label{calculationaldetails}

The calculations are performed within the density functional theory (DFT) \cite{DFT}
using the full potential augmented plane waves method, as implemented in the
Wien2k code \cite{Wien2k}. We use the local density approximation (LDA) \cite{LDA} for the 
exchange and correlation, and small muffin tin radii to allow for lattice 
relaxation. The muffin tin radii used were  R$^{MT}_{Ti}$ = 1.7 bohr,  R$^{MT}_{O}$ = 1.4 bohr,  R$^{MT}_{La}$ = 2.5 bohr,  R$^{MT}_{Al}$ = 1.7 bohr.  The 
number of plane waves in the interstitial region is mostly restricted
to RKMax=6 although in many cases it was increased to RKMax=7 and no 
qualitative difference was observed. The effect of increasing the number of
k-points was also checked. In some cases we relax the structure with the Siesta code \cite{SIESTAmethod}
as it allows for further cell deformation.

\section{TiO$_2$ anatase and LAO: Bulk and surface properties}
\label{bulkandsurface}

\begin{figure}[ht!]
\includegraphics[width=0.42\textwidth]{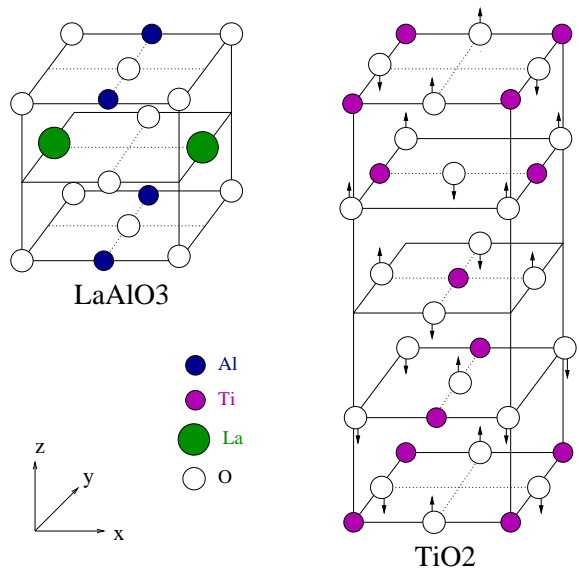}
\caption[]{\label{bulkunitcell} Bulk unit cells of the two component systems. The arrows on the oxygen atoms in the anatase structure indicate that they are off-plane with respect to the Ti atoms.}
\end{figure}

TiO$_2$ anatase and LAO are both conventional band insulators, with 
in-plane bulk lattice parameter  a=3.79\AA  \cite{TiO2_LAO_exp_latticeconst} (Fig. \ref{bulkunitcell}). The small mismatch between these in-plane lattice constants allows  for  a good epitaxial experimental outcome \cite{Lotnyk}.

\begin{figure}[ht!]
\includegraphics[width=0.4\textwidth]{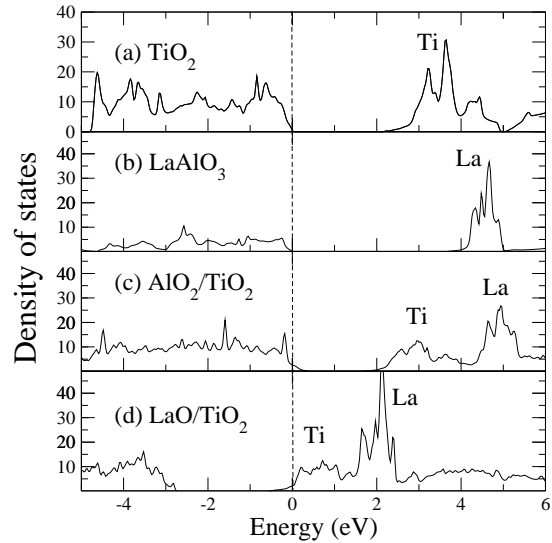}
\vspace{1cm}
\caption[]{\label{anatylao_dostot} Density of States of:  (a) bulk TiO$_2$ anatase (b) bulk LAO, (c) and (d) Superlattices either with AlO$_2$ or LaO planes in the interfaces (relaxed structures). Fermi energy is set to O (in all DOS graphs).}
\end{figure}

\begin{figure}[ht!]
\includegraphics[width=0.4\textwidth]{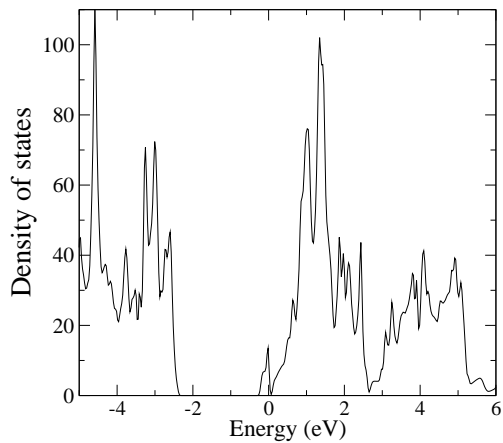}
\caption[]{\label{DOS_Tio2bulkwithvac} Density of states of bulk anatase with one oxygen vacancy (Ti$_{16}$ O$_{31}$). Notice the vacancy energy level very close to the conduction band.}
\end{figure}

TiO$_2$ anatase presents a 3.2 eV band gap between filled oxygen 2$p$ bands and 
unfilled Ti 3$d$ conduction bands . The calculated band gap is 2.2 eV as 
shown in Fig. \ref{anatylao_dostot}(a). When oxygen vacancies are formed in the bulk anatase 
structure the calculations show that the system becomes metallic and a vacancy 
level appears inside the gap, very close to the conduction band (Fig. \ref{DOS_Tio2bulkwithvac}). 
The small band gap and the vacancy levels very close to the conduction band 
are typical features of  the LDA approximation.

\vspace{1cm}

LAO presents a 5.6 eV gap  between filled oxygen 2$p$ bands hybridized with
Al $p$ states and unfilled conduction bands composed mostly of La 4$f$ states. The
calculated gap is 4 eV, as shown in Fig. \ref{anatylao_dostot}(b), and when oxygen vacancies are
created in this system the calculated vacancy levels appear well inside the 
band gap.

Summing up, both materials are non-magnetic insulating oxides in bulk and
if oxygen vacancies are present they keep their non-magnetic character although
the electronic structure changes, as vacancy levels appear inside the band gap.

\begin{figure}[ht!]
\includegraphics[width=0.35\textwidth]{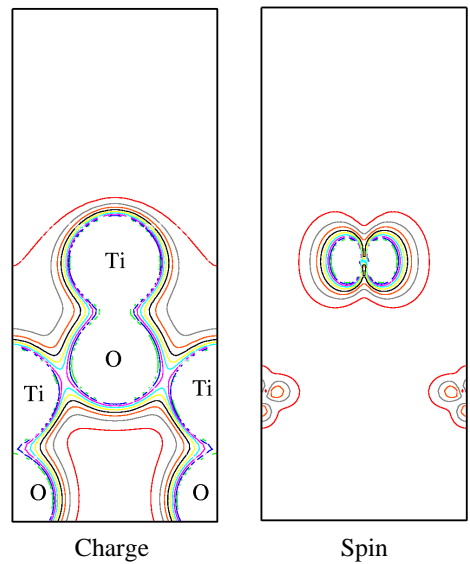}
\caption[]{\label{slabanatasa_CDSD} Charge and spin density of the surface of a TiO$_2$ slab with no oxygen atoms at the surface. Notice that only the surface  Ti atom has a considerable magnetic moment.}
\end{figure}

In order to get some insight about how these bulk properties  change
in the heterostructure, we study slabs of the component materials along the 
(001) direction.
The anatase slab with 5 or 9 TiO$_2$ layers is used 
to model the properties of the free surface of an anatase film. A different 
DFT calculation, using the SIESTA code, allows the unit cell to  change
its shape under relaxation, which is not the case with the Wien2k code. This 
results in a strong deformation of the (001) surface, that goes from a square 
into a rectangular shape, and also makes the slab thinner. The free standing 
anatase film would therefore deform in that way, but as we are studying its 
epitaxial growth over LAO we keep the in plane lattice constant fixed and 
the cell shape square. Relaxation is only considered  in the z direction,
perpendicular to the slab.
With this constraint we create oxygen vacancies in the surface
of the slab and relax the atomic positions using the Wien2k code. The
vacancies seem to be crucial for the appearance of magnetism: on  one
hand, magnetic solutions  appear if there are oxygen vacancies at the
surface and on the other hand the value of the magnetic moment increases
with the number of vacancies. In our calculations, one vacancy per surface 
(50\%) gives after relaxation a magnetic moment of 0.3 $\mu_B$, localized at 
the Ti atom in that surface, while two vacancies per surface give 2 $\mu_B$. 
In Fig. \ref{slabanatasa_CDSD} we show the charge and spin density maps near the surface for the  
anatase slab with two vacancies per surface (all the surface oxygen atoms removed).
The system is clearly magnetic, and the magnetic moment is mostly
localized in the Ti superficial atom. The effect of structural
relaxation is to move this Ti  atom closer to the remaining oxygen 
neighbor, which is apical and in the subsurface layer, thus making the slab 
thinner. In fact, this distance  decreases from 1.97 \AA \, in bulk TiO$_2$ to 
1.89 \AA \,  for a surface with no oxygen vacancies  and to 1.75 \AA \, if there 
are oxygen vacancies in the surface layer as in Fig. \ref{slabanatasa_CDSD}.
In  the slab with 5 TiO$_2$ layers  the two surfaces interact in such 
a way that the magnetic moments are smaller than in the 9-layer slab.

The other building block of the heterostructures is the slab of LAO.
Along the (001) direction this slab alternates layers of AlO$_2$ (negatively 
charged) and LaO (positively charged). Therefore,  an integer
number of unit  cells in the z direction would produce a large dipole and
it is more convenient to use an odd number of layers.
For the superlattices we consider 5 layer slabs with either LaO or AlO$_2$
termination. In both cases the system becomes metallic within the LDA
approximation and the interesting change with respect to the bulk density
of states, shown in Fig. \ref{anatylao_dostot}(b), is that the LaO termination shifts the 
unoccupied surface La states substantially towards the Fermi energy. This can
give as a result a large hybridization with the unoccupied Ti states of
TiO$_2$, but will not modify the occupied states.

\section{TiO$_2$/LAO superlattices}
\label{superlattices}

\begin{figure}[ht!]
\includegraphics[width=0.6\textwidth]{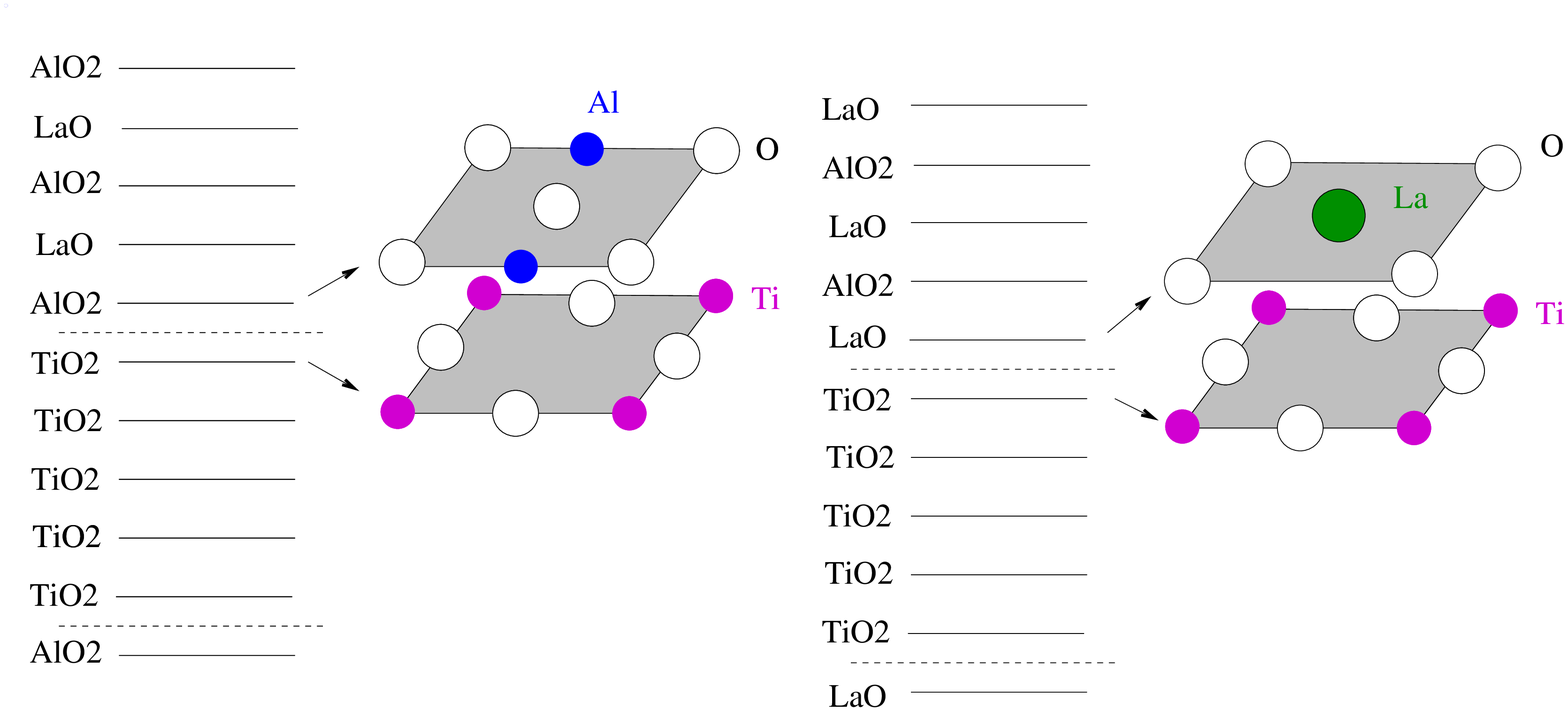}
\caption[]{\label{unitcell} Schematic diagrams for the unit cells of the superlattices studied.
Notice that there are two types of AlO$_2$/TiO$_2$ interfaces, due to the off-plane
oxygen atoms in the anatase structure (Fig. \ref{bulkunitcell}), but only one type of LaO/TiO$_2$
interface. }
\end{figure}

The system we propose to study is a TiO$_2$ (001) film with anatase structure
grown epitaxially over a LAO (001) substrate. For this purpose we perform 
calculations on the superlattices shown schematically in Fig. \ref{unitcell}. We have 
considered the 2 cases: the TiO$_2$ plane facing the AlO$_2$ or the  
LaO termination.  In both situations we assume that the oxygen atoms of 
the LAO surface face the Ti atoms of the anatase surface. The unit 
cell we use consists of 5 layers of each material and contains 27 or 28 atoms.
The lateral lattice parameter is kept fixed at the experimental value for
LAO but the distance between the two slabs composing the heterostructure
is obtained by minimizing the total energy. The fact that the total energy
has a minimum when changing this distance  verifies that there  is bonding 
between the two materials and also gives  the optimal value
for that distance. This fixes the size of the superlattice unit cell along
the z axis for each case, for example we notice that it changes when there 
are oxygen vacancies. Afterwards, all the 
atoms are allowed to relax inside the unit cell until the force on each one 
is less than 2 meV/\AA.

We find that there is a competition between the tendency of the anatase film
to shrink, as in the free standing slab, and the tendency to bond to LAO.
In particular, if an oxygen from the anatase surface faces an Al atom the
bond is strong. Also, 
there is an electrostatic effect due to the charge in each layer of LAO,
which is of course different for a neutral TiO$_2$ layer and for a charged
layer, when oxygen vacancies are created in the anatase surface.

Fig. \ref{anatylao_dostot}(c) and (d) show the density of states of the two superlattices without vacancies and after relaxation. These graphs give some insight about electron transfer or the possibility of oxygen vacancy formation, in agreement with experiments \cite{NakagawaNatMat06}. In the case of the AlO$_2$/TiO$_2$ interface, the states readily available near E$_F$ are O-2$p$ states. This interface will then be prone to the formation of oxygen vacancies in order to allow for a compensation of the polar discontinuity present at the interface. For the  LaO/TiO$_2$ interface instead, the states near E$_F$ are mainly comprised of Ti 3$d$-states.  This  accessibility to a mixed valency of Ti  allows for a transference of electrons across the interface to overcome the electrostatic potential divergence.  Both interfaces are metallic without vacancies. This metallicity is not altered by the formation of oxygen vacancies or by structural relaxation. 

\subsection{AlO$_2$/TiO$_2$ heterointerface}

\begin{figure}[ht!]
\includegraphics[width=0.35\textwidth]{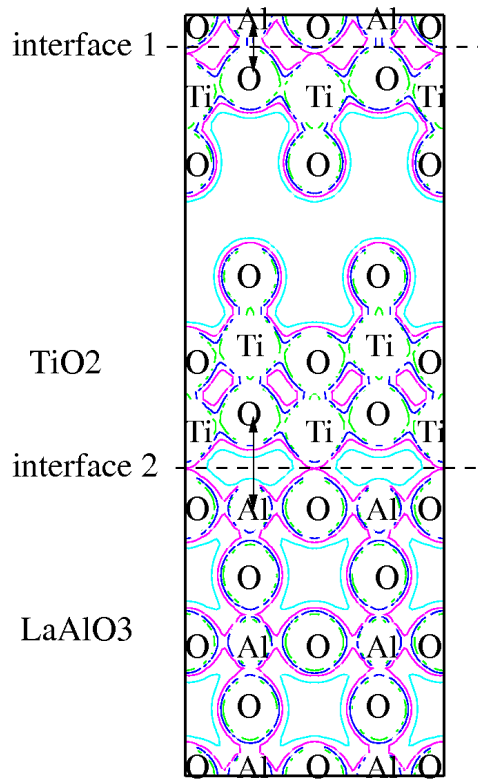}
\caption[]{\label{anatylao_interfaces} Charge density for the superlattice with AlO$_2$/TiO$_2$ interfaces and no vacancies. The two interfaces are different: one can see that the distance from the surface Al to the neighbor anatase oxygen is shorter in interface 1.}
\end{figure}

Due to the anatase structure there are two types of  AlO$_2$/TiO$_2$ interfaces. This can be inferred from Fig. \ref{bulkunitcell} and 
Fig. \ref{unitcell}(a), but it is shown more clearly in Fig. \ref{anatylao_interfaces}.  To simulate growth conditions in a typical experiment \cite{BrinkmanNatMat07}, we introduced oxygen vacancies at both layers of the interface and found that it  is more favorable to perform them in the TiO$_2$ layer rather than in the AlO$_2$ plane.

When the heterostructure is formed, the  interface becomes metallic; with no oxygen vacancies the Fermi level lies 
in the valence band (Fig. \ref{anatylao_dostot}(c)) and with vacancies it moves to the conduction band. 

\begin{figure}[ht!]
\includegraphics[width=0.5\textwidth]{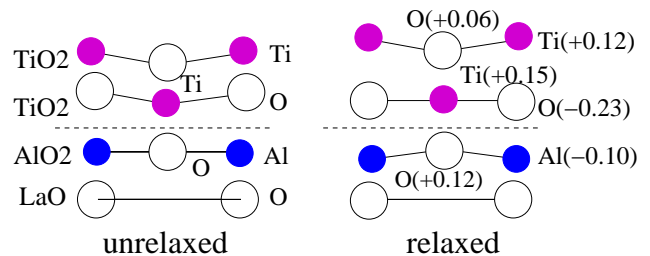}
\caption[]{\label{relaxation} Schematic diagram showing the atomic relaxation on the magnetic interface (namely, interface 2 in Fig. \ref{anatylao_interfaces}) when there is one oxygen vacancy at the anatase surface. The movement of each atom (in \AA) is indicated  when it is significant (more than 0.05 \AA) }
\end{figure}

We find magnetic solutions in the unrelaxed geometry, with and without
vacancies, but the magnetic moments decrease with  structural relaxation
as can be seen in Table 1.  Without vacancies the moments are located in some
of the oxygen atoms and disappear with relaxation. With vacancies they are 
located in the Ti superficial atoms and do not disappear, they only decrease 
somewhat. One particular example of the structural relaxation is shown in 
Fig. \ref{relaxation}, indicating schematically how each of the atoms moves in the magnetic 
interface. The atom that has a larger displacement is the anatase oxygen 
facing Al,  it bonds more strongly with Al than with Ti. Relaxation changes the buckling of the anatase surface.

\begin{figure}[ht!]
\includegraphics[width=0.45\textwidth]{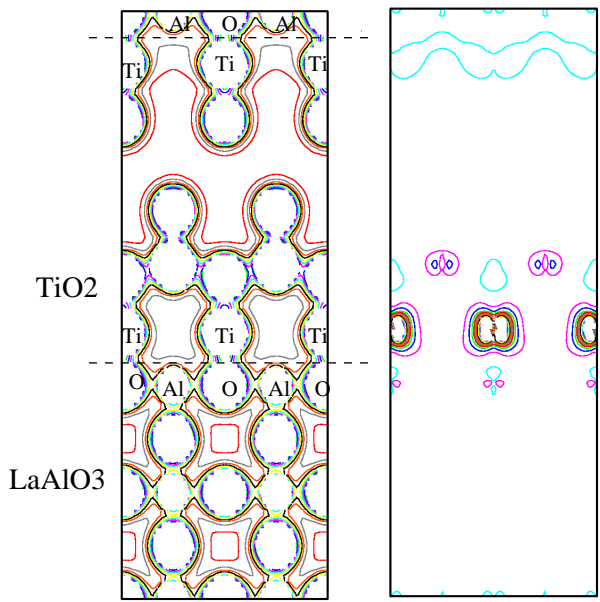}
\caption[]{\label{anatylao_CDSD_sinmini} Charge and spin density maps for the superlattice with AlO$_2$/TiO$_2$ interfaces and 2 vacancies.  The effect of relaxation is too small to be seen in this scale. Notice that only one of the interfaces is magnetic.}
\end{figure}

Fig. \ref{anatylao_CDSD_sinmini} shows the charge and spin densities of the superlattice, when
there are no oxygen atoms in the TiO$_2$ side of the interfaces (2 vacancies).
 Although both interfaces have the same number of missing oxygen atoms and first neighbours,
only one of them is magnetic, depending on the relative orientation of 
the second neighbor layers. This points to the new result that magnetism is not only due to interactions between interfacial atoms but also further neighbours are important.
 
\begin{figure}[ht!]
\includegraphics[width=0.42\textwidth]{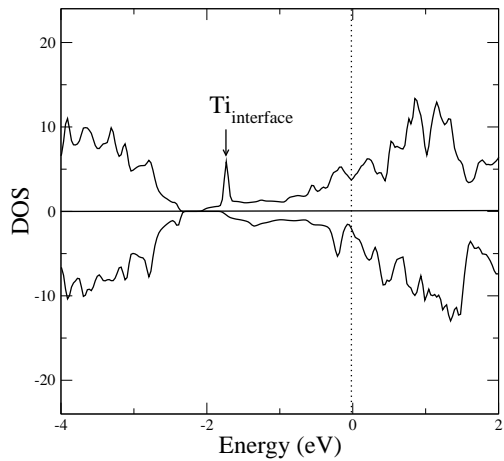}
\caption[]{\label{DOSanatylao2vac}  Density of states for the case of Fig. \ref{anatylao_CDSD_sinmini} with the relaxed structure. The states corresponding to the Ti at interface 2 are highlighted. }
\end{figure}

The corresponding density of states is shown in Fig. \ref{DOSanatylao2vac} where the Ti 3$d$ states are marked and they are mainly composed of d$_{x2-y2}$ and d$_{xy}$ states.

\subsection{LaO/TiO$_2$  heterointerface}

To study this heterointerface we use the supercell depicted in Fig. \ref{unitcell}(b) and perform oxygen vacancies to simulate experimental growth conditions.  It is more favorable to perform them in the TiO$_2$ layer compared to the LaO face. Introducing vacancies in this case is overall less favourable compared to the other interface and  magnetic solutions are only found with many oxygen vacancies present, as shown in Table 1. 
As in this case both interfaces of the superlattice are identical, the magnetic moment per Ti atom at the surface is considerably smaller than in the previous interface.
In Fig. \ref{anatylao_dostot}(d) we can see that  hybridization of La 4$f$  with Ti 3$d$ electrons is large, probably exaggerated by the LDA approximation, as was mentioned in Ref. \cite{MillisSpaldin_PRL06}.  Comparing the valence electron charge inside the muffin-tin spheres for Ti with those of their counterparts in bulk, we corroborate a mixed valency state for the Ti atoms at the interface without vacancies while in the previous superlattice it was due to the vacancies. 

Atomic relaxation in this interface is considerably smaller than in the previous one.

\subsection{Cohesive energies}
\label{Cohesivenergies}

As a very first approximation one could estimate the cohesive energies of
the systems in their relaxed structures by subtracting the energy
of the conforming atoms from the total energy of the system. By doing so 
we obtain that a 5 layer slab of LAO prefers the termination AlO$_2$ 
over LaO. The cohesive energy of a superlattice with AlO$_2$/TiO$_2$ 
interfaces is also larger than that with LaO/TiO$_2$ interfaces.

A better approximation to the cohesive energy of a superlattice, given in 
Table 1, is to subtract from its total energy the energy of the relaxed slabs 
of TiO$_2$ and LAO that compose it. This gives a more interesting result: the supercell with LaO/TiO$_2$ interface has more 
cohesive energy if no vacancies are present while the one with AlO$_2$/TiO$_2$ 
interface is preferred if there are oxygen vacancies. This last case presents 
a magnetic moment, localized at the superficial Ti atom. It is important to note that the energy difference between cases with one and two vacancies is order of magnitude smaller compared to the cases with no vacancies. Therefore, with our results we can draw qualitative conclusions comparing cases either with or without vacancies but not referring to the concentration of vacancies.

\begin{table}[h]
\begin{center}\begin{tabular}{|c|ccc|ccc|}
\hline
 case & &Ti/Al &&& Ti/La&  \\
\hline
   & E (eV) && $\mu$ ($\mu_B$) &  E(eV) & & $\mu$  ($\mu_B$) \\
\hline
\hline
no vac &   4.89 &&0.75 $\rightarrow$ 0.01  &   6.86 && 0  \\
\hline
1 vac   &  6.65  &&0.38 $\rightarrow$ 0.26&   6.01&& 0 \\
\hline
2 vac    &  6.62 && 1.36$\rightarrow$ 1.28 &   5.94 & &1.05$\rightarrow$ 0.57  \\
\hline
\end{tabular}\end{center}
\label{table}
\caption{Cohesive energies and total magnetic moments per unit cell for the two types of superlattices considered, with or without vacancies (vac). The effect of lattice relaxation is indicated by the arrows for the magnetic moments. The number of vacancies is the number of missing oxygen atoms from the anatase side of each interface.}
\end{table}

\section{Conclusions}
\label{conclusions}

Our calculation shows that magnetism appears at the TiO$_2$ anatase surface and at the TiO$_2$/AlO$_2$ interface when there are enough oxygen vacancies. Structural relaxation is found to be significant and produces an important effect on the superficial Ti magnetic moment, diminishing it. 
The experimental substrate LAO,  presents  large terraces in its surface but it is not clear which interface predominates. Using energetic arguments we find that TiO$_2$ deposition over both LAO layers could occur with equal probability but with different oxygen content in each case. The interface with AlO$_2$ has the largest cohesive energy when there are oxygen vacancies, while when there are no vacancies, the interface with LaO is preferred. Therefore, if the more abundant terraces have LaO termination, the  magnetism observed experimentally would not be interfacial. It could be due to vacancies in the free surface of the anatase film, and the  role of the substrate would be only to fix the cell size. If on the contrary the more abundant interface would be with Al, and the system is grown under low oxygen pressure, there could be a magnetic moment localized at the interface.

\begin{acknowledgments}
We acknowledge discussions with L. Errico and R. Weht. This work was funded by CONICET-Argentina.  MW and VF are members of CIC-CONICET.
\end{acknowledgments}


\end{document}